\documentclass[sigconf, nonacm]{acmart}
\AtBeginDocument{%
  }

\usepackage{multirow}
\usepackage[ruled,vlined]{algorithm2e}
\usepackage{enumitem}
\usepackage{pifont}
\usepackage{soul}
\usepackage[table]{xcolor}
\usepackage{subfigure}
\usepackage{balance}

\newcommand{\name}[0]{DynaTree\xspace}

\copyrightyear{2026}
\acmYear{2026}
\setcopyright{cc}
\setcctype{by-nc-nd}
\acmConference[KDD '26]{Proceedings of the 32nd ACM SIGKDD Conference on Knowledge Discovery and Data Mining V.2}{August 09--13, 2026}{Jeju Island, Republic of Korea}
\acmBooktitle{Proceedings of the 32nd ACM SIGKDD Conference on Knowledge Discovery and Data Mining V.2 (KDD '26), August 09--13, 2026, Jeju Island, Republic of Korea}
\acmDOI{10.1145/3770855.3818386}
\acmISBN{979-8-4007-2259-2/2026/08}

\acmSubmissionID{v2ads498}

\begin{document}

\title{DynaTree: Dynamic Agentic Retrieval Tree for Time-Sensitive News Retrieval}

\author{Siyuan Qi}
\orcid{0009-0009-3445-918X}
\affiliation{%
  \institution{Shanghai Jiao Tong University}
  \city{Shanghai}
  \country{China}
}
\email{qisiyuan7936@sjtu.edu.cn}

\author{Xinyuan Wang}
\affiliation{%
  \institution{Shanghai Jiao Tong University}
  \city{Shanghai}
  \country{China}
}
\email{wangxinyuan1@sjtu.edu.cn}

\author{Yingxuan Yang}
\affiliation{%
  \institution{Shanghai Jiao Tong University}
  \city{Shanghai}
  \country{China}
}
\email{zoeyyx@sjtu.edu.cn}

\author{Haochuan Guo}
\affiliation{%
  \institution{Orion Arm AI}
  \city{Singapore}
  \country{Singapore}
}
\email{haochuan.guo@orionarm.ai}

\author{Jianghao Lin}
\affiliation{%
  \institution{Shanghai Jiao Tong University}
  \city{Shanghai}
  \country{China}
}
\email{linjianghao@sjtu.edu.cn}

\author{Weiwen Liu}
\authornote{Corresponding author.}
\affiliation{%
  \institution{Shanghai Jiao Tong University}
  \city{Shanghai}
  \country{China}
}
\email{wwliu@sjtu.edu.cn}

\author{Yong Yu}
\affiliation{%
  \institution{Shanghai Jiao Tong University}
  \city{Shanghai}
  \country{China}
}
\email{yyu@apex.sjtu.edu.cn}

\author{Weinan Zhang}
\authornotemark[1]
\affiliation{%
  \institution{Shanghai Jiao Tong University}
  \city{Shanghai}
  \country{China}
}
\email{wnzhang@sjtu.edu.cn}

\renewcommand{\shortauthors}{Qi et al.}

\begin{abstract}
  Agentic Retrieval-Augmented Generation (RAG) improves retrieval
by integrating planning, tool use, and iterative reasoning,
but most existing agentic RAG methods tightly couple
semantic expansion with retrieval decisions
within short-horizon inference loops,
resulting in high inference costs
and limited suitability for time-sensitive, high-recall tasks
such as news retrieval.
We propose \textbf{DynaTree}
(\textbf{Dyna}mic Agentic Retrieval \textbf{Tree}),
a two-stage framework in which
agentic RAG is used for offline semantic exploration,
while online retrieval is performed without agentic inference.
In the first stage,
DynaTree adopts an agentic RAG process
to construct a reusable retrieval tree for each query topic,
where coordinated agents plan semantic expansions,
execute diverse retrieval strategies,
normalize retrieved evidence, and iteratively refine the tree using retrieval feedback,
materializing a persistent representation
of the query’s semantic space.
In the second stage,
DynaTree operates purely as a structure-aware retrieval system,
performing lightweight daily subtree selection
over a time-localized evaluation proxy, 
with large language models used only for rapid relevance labeling,
without further agentic reasoning,
tree modification, or retraining.
Experiments on our multi-day Syft news benchmark
and multiple BEIR datasets show that DynaTree
achieves state-of-the-art recall and NDCG,
consistently outperforming standard RAG
and prior agentic baselines.
Moreover, DynaTree has been directly deployed
in the Syft production system and evaluated
through online A/B testing from Jan.~28 to Feb.~6,~2026.
Across all days, the dynamically adapted variant
with daily subtree selection improves survival rate
from $0.32$--$0.53$ to $0.59$--$0.73$,
yielding consistent gains of up to $\sim\!1.5\times$
over a fixed offline-selected subtree
and outperforming all existing production recallers
on every evaluation day.
These online gains persist under live user traffic,
demonstrating that DynaTree translates
structure-aware, agentic offline reasoning
into tangible improvements in coverage, freshness,
and relevance for real-world news retrieval.
\end{abstract}




\maketitle

\section{Introduction}
\label{sec:introduction}

\begin{figure}[t]
    \centering
    \subfigure{
        \includegraphics[width=0.3\linewidth]{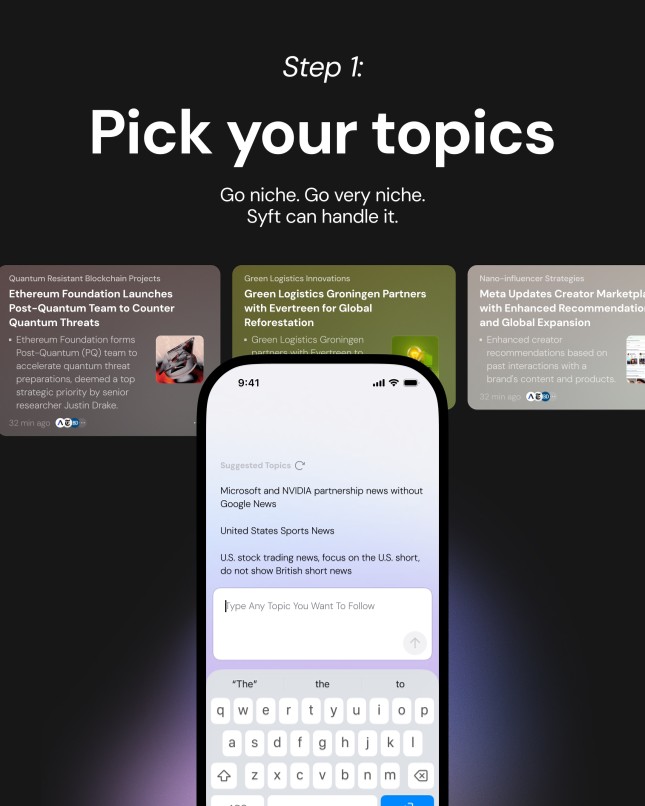}
    }
    \subfigure{
        \includegraphics[width=0.3\linewidth]{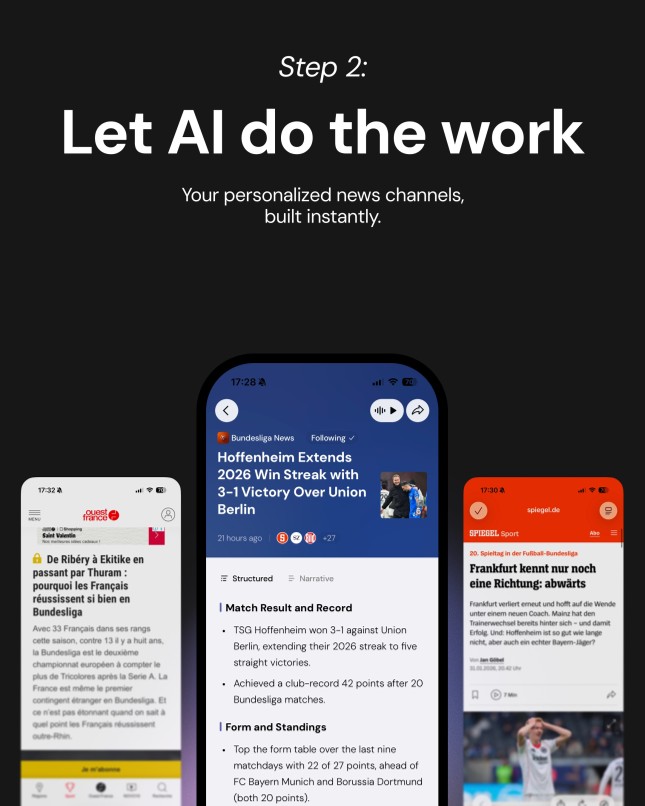}
    }
    \subfigure{
        \includegraphics[width=0.3\linewidth]{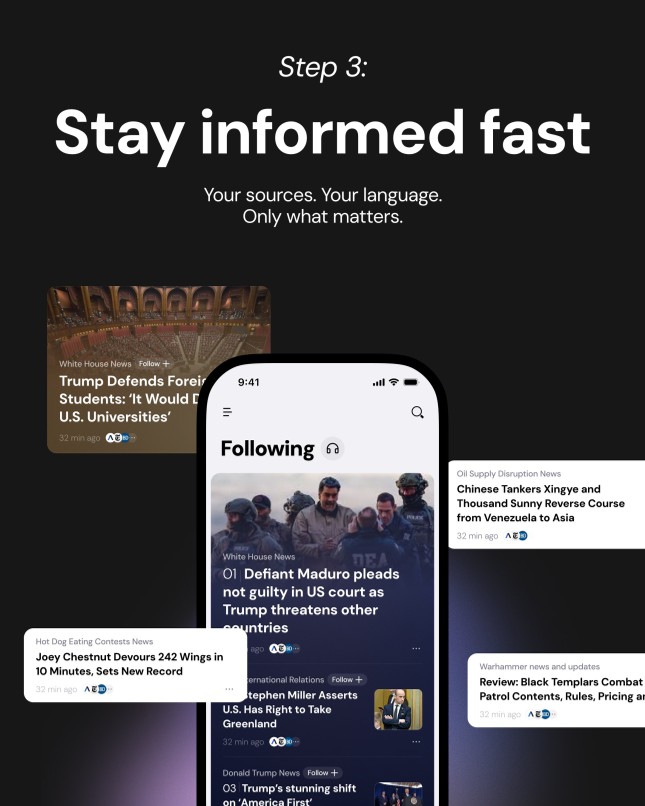}
    }
    \caption{Syft product showcase. Our retrieval strategy is deployed in live traffic, delivering a better user experience.}
    \label{fig:four_inline}
\end{figure}

News retrieval aims to identify a comprehensive set of 
relevant articles for downstream ranking, recommendation, and content analysis in modern news systems~\cite{tang-etal-2025-uncovering}. 
However, user queries in this domain are often abstract or underspecified, particularly when addressing emerging events, breaking stories, or evolving topics, making effective semantic query expansion a critical prerequisite for high-recall retrieval~\cite{10.1145/2071389.2071390,10.1145/383952.383972,li2026queryexpansionagepretrained}. Moreover, news corpora exhibit pronounced temporal dynamics: new articles are continuously published while previously relevant articles quickly become outdated, leading to frequent and abrupt shifts in relevance distributions~\cite{10.1145/956863.956951,10.1145/1718487.1718489}. These characteristics demand retrieval mechanisms that are both \textit{semantically expressive} and \textit{temporally adaptive}.

Traditional news retrieval methods address semantic sparsity and temporal drift through query expansion~\cite{10.1145/2071389.2071390}
and freshness-aware ranking~\cite{piryani2026itshightimesurvey}. 
Techniques such as pseudo-relevance feedback and language-model-based expansion enrich under-specified queries to improve recall~\cite{10.1145/383952.383972}, while complementary approaches leverage temporal signals to prioritize recent or temporally aligned content~\cite{10.1145/956863.956951,10.1145/1718487.1718489}. However, these methods typically perform expansion in an ad hoc, per-query manner and lack persistent semantic structures that evolve over time. This limits their ability to balance effectiveness and rapid adaptation to shifting relevance in real-world settings.

Recently, retrieval-augmented generation (RAG) has become a standard paradigm
for grounding language models in external knowledge, improving semantic coverage
in knowledge-intensive and time-sensitive information access settings~\cite{10.5555/3495724.3496517,izacard-grave-2021-leveraging,pmlr-v162-borgeaud22a,vu-etal-2024-freshllms}. Agentic extensions to classical RAG introduce planning, iterative query reformulation, and self-refinement, fostering more expressive and adaptive semantic expansion~\cite{yao2023reactsynergizingreasoningacting,10.5555/3666122.3666499,ICLR2024_25f7be96,10.5555/3692070.3694642}. While effective for complex reasoning, many agentic RAG approaches operate via short-horizon, per-query inference loops, incurring significant latency and token costs through redundant, on-the-fly expansions. This tight coupling of reasoning and retrieval hinders reuse across queries or time, limiting their viability in time-sensitive, efficient news settings~\cite{Wang_2024, singh2026agenticretrievalaugmentedgenerationsurvey}.

A fundamental limitation of existing agentic RAG systems is that semantic expansion is treated as a transient byproduct of per-query reasoning rather than a persistent asset. This design fails to capitalize on two critical opportunities: reuse and relational modeling. Because expansions are discarded after each request, the system suffers from redundant computation and cannot accumulate semantic knowledge over time. Consequently, it fails to model how subtopics relate to one another or how a query's facets evolve. This is particularly detrimental in news retrieval, where high-level intent (e.g., "AI in consumer electronics") remains stable, but specific subtopics and surface realizations shift daily. Current methods tend to overfit recently salient terms (e.g., "AI chips") while neglecting semantically coherent but less prominent subtopics like "on-device inference" or "privacy-preserving personalization." Over time, this structural mismatch between transient expansions and persistent intent causes systematic recall loss for relevant but underexplored content.

Motivated by this observation, we argue that semantic expansion must be materialized as a persistent, structured representation that captures the relationships between topics, rather than serving as a transient artifact. While prior graph-based methods improve matching efficiency~\cite{edge2025localglobalgraphrag,zhu-etal-2025-knowledge}, they fail to address how such structures can be constructed via agentic reasoning or adapted to temporal dynamics. To bridge this gap, we propose \textbf{\name}, a framework that decouples expansion from real-time retrieval to enable reuse. In the first stage, we amortize agentic RAG to build a reusable retrieval tree that maps the query's semantic space. In the second stage, a lightweight daily subtree selection mechanism adapts to evolving news distributions without repeated agent execution or model retraining. This design achieves high recall with low latency, making \name ideal for real-time retrieval over continuously updated corpora. 
Finally, our method has been deployed
in the production-level news retrieval system \textbf{Syft}
and evaluated on real user traffic through online A/B testing.
Over a ten-day period (Jan.~28--Feb.~6,~2026),
the dynamically adapted variant with daily subtree selection
consistently improves survival rate
from $0.32$--$0.53$ to $0.59$--$0.73$,
yielding up to $\sim\!1.5\times$ gains
over a fixed offline-selected subtree
and outperforming all existing production recallers. 
Figure~\ref{fig:four_inline} provides an overview
of the deployed product pipeline.

Our main contributions are summarized as follows:
\begin{itemize}[leftmargin=*, itemsep=2pt, topsep=2pt]
\item We propose \name, the first agentic framework to decouple semantic expansion from online retrieval. By materializing topic relationships into a reusable retrieval tree and enabling adaptation via lightweight subtree selection, it effectively reconciles the trade-off between recall, adaptability, and latency in real-time news retrieval.
\item We introduce the Syft News dataset, a large-scale benchmark designed to evaluate high-recall retrieval under realistic temporal distribution shifts. Extensive experiments on this and public BEIR benchmarks demonstrate that \name consistently outperforms strong RAG and agentic baselines.
\item We report on the successful deployment of \name within the Syft news platform's production pipeline. Operating on live traffic under strict latency constraints, the system delivered measurable gains in retrieval quality, validating its practical effectiveness.
\end{itemize}

\section{Related Work}
\label{sec:related}

\paragraph{Retrieval-Augmented Generation and Agentic Retrieval}

Retrieval-Augmented Generation (RAG) grounds language model outputs
in externally retrieved evidence, improving factuality and coverage
for knowledge-intensive tasks~\cite{10.5555/3495724.3496517}.
Subsequent work has systematized RAG design choices,
including retriever--generator coupling,
context selection,
and efficiency trade-offs ~\cite{gao2024retrievalaugmentedgenerationlargelanguage},
while recent surveys on LLMs for information retrieval (LLM4IR)
highlight the role of large language models
as query rewriters, rerankers, and reasoning modules
within retrieval pipelines~\cite{10.1145/3748304}.
A key emerging direction is \emph{agentic retrieval},
where models plan,
invoke tools,
and iteratively refine retrieval through reasoning steps.
Representative frameworks such as ReAct and Reflexion
demonstrate how step-wise tool use and self-reflection
can improve retrieval-informed generation
~\cite{yao2023reactsynergizingreasoningacting,10.5555/3666122.3666499},
while complementary work emphasizes query rewriting
as a structured interface between LLM reasoning and retrieval,
showing that rewritten queries can better align retrieval
with downstream generation objectives~\cite{ma-etal-2023-query}.
More recent agentic RAG systems incorporate
self-evaluation and adaptive retrieval control
to improve robustness across queries~\cite{ICLR2024_25f7be96,yao-etal-2025-seakr}.
Despite these advances,
most existing approaches tightly couple
semantic expansion, retrieval, and reasoning
within per-query or short-horizon inference loops,
making semantic expansions transient
and computationally expensive to repeat
in continuously updated environments.
In contrast,
our framework materializes semantic expansion
as a persistent \emph{retrieval tree}
and decouples \emph{offline semantic exploration}
from \emph{daily lightweight optimization},
enabling amortized agentic reasoning
and structural reuse for time-sensitive news retrieval.

\paragraph{Semantic Expansion and Time-Sensitive Information Retrieval}

Semantic expansion has been a central technique
for improving recall on abstract or underspecified queries.
Classic relevance feedback and probabilistic relevance models
provide principled ways to enrich query representations
using pseudo-relevant documents~\cite{10.1145/383952.383972}.
Subsequent surveys summarize pseudo-relevance feedback
and knowledge-based expansion
as core paradigms in information retrieval
~\cite{10.1145/2071389.2071390,10.1016/j.ipm.2019.05.009}.
In parallel,
time-sensitive retrieval introduces additional challenges,
as relevance distributions shift
with the continuous arrival of new documents
and the rapid obsolescence of older information.
Prior temporal IR work
has studied recency-aware ranking
and temporal signals for web and news search
~\cite{10.1145/1772690.1772725,10.1145/1718487.1718490},
as well as broader formulations of temporal information retrieval
and benchmark tasks
~\cite{10.1145/2911451.2914805,DBLP:conf/trec/AslamEPDMS14}.
Classic stream-based settings, such as knowledge-base acceleration
and temporal summarization, study timely updates or meaningful deltas
from incoming document streams~\cite{DBLP:conf/trec/FrankBKRTZRVS13,DBLP:conf/trec/AslamEPDMS14}.
They are relevant to time-sensitive retrieval,
but are often entity-centric or update-centric.
More recently,
structured and graph-based retrieval methods
have been proposed to better capture semantic relationships
among documents or concepts,
including graph-augmented RAG
and hierarchical or tree-organized retrieval schemes~\cite{edge2025localglobalgraphrag,DBLP:conf/iclr/SarthiATKGM24,zhu-etal-2025-knowledge}.
While such approaches improve global semantic coverage
or multi-hop interaction,
their structures are typically constructed
from static corpora or document relations
and are not explicitly designed
to be reused or adapted under strong temporal dynamics.
Recent LLM-based methods further address temporal gaps
by conditioning generation on freshly retrieved documents
or incorporating time awareness into model behavior,
but adaptation is still performed
at the document or query level
without maintaining a persistent semantic structure across time~\cite{vu-etal-2024-freshllms,piryani2026itshightimesurvey}.
In contrast, our approach explicitly separates \emph{stable semantic coverage} from \emph{daily temporal adaptation}: the former is materialized as a reusable agentic retrieval tree, while the latter is implemented through lightweight subtree selection over a time-localized proxy. This design enables high recall in dynamic news environments at substantially reduced cost.

\section{Methodology}
\label{sec:methodology}

\begin{figure*}[t]
    \centering
    \includegraphics[width=\textwidth]{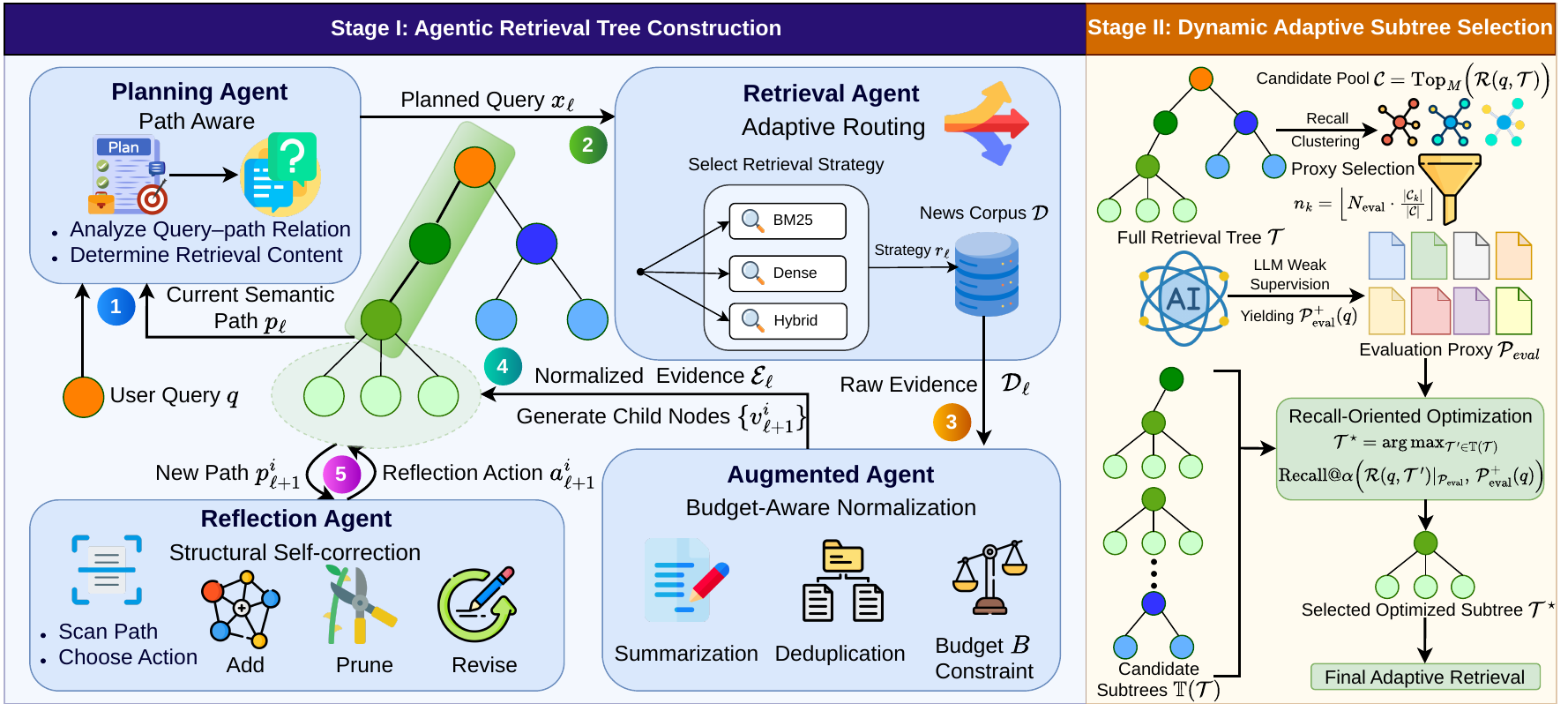}
    \caption{Overview of our framework DynaTree.
Stage I constructs a reusable retrieval tree through role-specialized agent collaboration, integrating path-aware planning, adaptive retrieval, budget-aware normalization, and structural self-correction.
Stage II dynamically selects the optimized subtree from pre-built tree via a lightweight evaluation proxy and recall-oriented optimization, enabling adaptive retrieval under evolving news distributions without re-running semantic expansion.}
    \label{fig:framework}
\end{figure*}

\subsection{Problem Formulation and Framework}
\label{subsec:problem}

We study the problem of \emph{semantic expansion for effective news retrieval}
through an explicit \emph{retrieval tree} abstraction.
Given a user query $q$ and a large news corpus
$\mathcal{D}=\{d_1,\dots,d_N\}$,
our goal is to systematically explore the semantic space induced by $q$,
materialize this exploration as a reusable and structured representation,
and leverage it to improve retrieval quality
under realistic ranking cutoffs.
Most existing query expansion and agentic RAG approaches rely on flat or short-horizon expansions,
which tightly couple semantic exploration with per-query retrieval decisions
and hinder structural reuse under evolving news distributions. In contrast, we model semantic exploration as a \emph{tree-structured process},
where different branches capture distinct subtopics,
semantic interpretations,
and refinement trajectories of the original query.
By explicitly organizing semantic expansion into a structured retrieval tree,
our abstraction enables systematic comparison, pruning, and reuse
of alternative retrieval behaviors,
and provides a well-defined foundation
for adaptive retrieval under dynamic news environments.

\paragraph{Retrieval Tree and Subtree-Induced Retrieval.}
Formally, a retrieval tree $\mathcal{T}$ is rooted at the original query $q$.
Each node represents a semantic expansion,
and edges encode refinement or specialization relations between expansions.
A root-to-leaf path
\begin{equation}
p = (v_0, v_1, \dots, v_L), \quad v_0 = q
\end{equation}
defines a coherent semantic trajectory
that progressively refines the original query.
Rather than treating individual expansion terms independently,
we regard each path as an \emph{atomic semantic unit}
that captures a complete, structured interpretation of the query,
preserving both subtopic context and refinement order.

While the full tree $\mathcal{T}$ provides broad semantic coverage,
using all paths simultaneously may introduce redundancy
or amplify semantically misaligned expansions.
We therefore consider a feasible set of candidate subtrees
$\mathbb{T}(\mathcal{T})$,
where each subtree $\mathcal{T}' \subseteq \mathcal{T}$
is a connected subgraph containing the root.
Each subtree induces a distinct retrieval behavior
through the collection of semantic paths it retains.
We denote by $\mathcal{R}(q, \mathcal{T}')$
the ranked list of documents returned by the retrieval-and-ranking procedure induced by $\mathcal{T}'$.

Given a relevance set $\mathcal{D}^{+}(q) \subseteq \mathcal{D}$, objective is to select the subtree
that maximizes retrieval effectiveness.
We instantiate this using Recall
to emphasize semantic coverage at retrieval stage:
\begin{equation}
\mathcal{T}^{\star}
=
\arg\max_{\mathcal{T}' \in \mathbb{T}(\mathcal{T})}
\;\mathrm{Recall}\bigl(
\mathcal{R}(q, \mathcal{T}'),\;
\mathcal{D}^{+}(q)
\bigr)
\label{eq:objective}
\end{equation}

We use Recall as the selection objective
to emphasize semantic coverage at the retrieval stage,
and report ranking-oriented metrics such as NDCG@10 in evaluation.
This formulation explicitly decouples
\emph{semantic exploration} from \emph{retrieval optimization}:
the retrieval tree is constructed once
to comprehensively capture the structured semantic space of a query (Stage~I),
while retrieval behavior is adaptively optimized over time
via lightweight subtree selection (Stage~II).

\paragraph{Path-Based Encoding and Scoring.}
Each subtree $\mathcal{T}'$ induces a set of root-to-leaf paths
$\mathcal{P}(\mathcal{T}')$,
which we interpret as an ensemble of complementary
semantic trajectories covering different subtopics
and refinement perspectives of the query.
For each path $p \in \mathcal{P}(\mathcal{T}')$,
we construct a unified textual representation $x_p$
by composing the textual descriptions
along the refinement trajectory,
and encode it holistically using a text-embedding model:
\begin{equation}
\mathbf{e}_p
=
\mathrm{Enc}(x_p)
\in \mathbb{R}^{d}
\label{eq:path_emb_stage3}
\end{equation}

Given a document $d$ with embedding $\mathbf{e}_d$,
its relevance score under subtree $\mathcal{T}'$
is computed by aggregating semantic similarity
across all induced paths:
\begin{equation}
s(d \mid \mathcal{T}')
=
\frac{1}{|\mathcal{P}(\mathcal{T}')|}
\sum_{p \in \mathcal{P}(\mathcal{T}')}
\cos\!\left(
\mathbf{e}_d,\;
\mathbf{e}_p
\right)
\label{eq:path_ensemble_score}
\end{equation}
This path-ensemble formulation favors documents
that receive consistent semantic support
from multiple structured trajectories,
thereby promoting comprehensive semantic coverage
in recall-oriented news retrieval settings.

\paragraph{Discussion: Why a tree structure?}
DynaTree adopts a tree because Stage~I expansion is inherently path-conditioned and recursive: each node is generated along a specific root-to-node semantic trajectory, making each path an ordered expansion unit. This structure also makes Stage~II connected-subtree selection tractable under a fixed online budget. In contrast, general graphs introduce path explosion and unstable structure selection. Thus, the tree is not only a representation choice, but also a computational constraint for efficient latency-sensitive adaptation.

\subsection{Agentic Retrieval Tree Construction}
\label{subsec:stage1}

Stage~I constructs a retrieval tree $\mathcal{T}$ that systematically explores semantic space of a query for effective news retrieval,
while explicitly controlling redundancy, semantic drift,
and unproductive expansion.
Rather than treating retrieval-augmented generation
as a monolithic, per-query operation,
we formulate tree construction as a
\emph{role-specialized agentic process}
that decomposes semantic exploration into planning, retrieval,
evidence normalization, and structural self-correction.
This decomposition enables fine-grained control and reasoning
over different aspects of semantic expansion within an agentic RAG framework,
resulting in a structured, interpretable,
and reusable representation that captures diverse query interpretations
and their underlying subtopic structure.

\paragraph{Recursive Expansion as a Controlled Operator.}
Let $p_\ell = (v_0,\dots,v_\ell)$ denote a root-to-node path
in the retrieval tree.
Each expansion step applies a controlled expansion operator
that maps the current path
to a set of semantically distinct child nodes:
\begin{equation}
\{v_{\ell+1}^{i}\}
=
\mathcal{F}_{\mathrm{agent}}
\bigl(q,\; p_\ell,\; \mathcal{D}\bigr)
\label{eq:agentic_expand}
\end{equation}
where $\mathcal{F}_{\mathrm{agent}}$ denotes a coordinated interaction
among agents,
subject to explicit structural constraints
such as max tree depth, branching factor,
and semantic diversity.
This abstraction enables high-level control over growth and coverage of the retrieval tree,
while remaining agnostic to internal realization
of individual agents.

\paragraph{Planning Agent: Path-aware Semantic Planning.}
The Planning Agent proposes a retrieval query
jointly conditioned on the original user query $q$
and the semantic trajectory encoded by the current path $p_\ell$.
This process is modeled as a stochastic policy:
\begin{equation}
x_\ell \sim \pi_{\mathrm{plan}}(q, p_\ell)
\end{equation}
where $x_\ell$ denotes the planned retrieval query at depth $\ell$.

By conditioning on the entire root-to-node path
rather than the current node alone,
the Planning Agent explicitly incorporates expansion history
and accumulated semantic context.
This path-aware design encourages \emph{semantic progression}
along the retrieval tree,
discouraging shallow paraphrasing
and redundant reformulations.
Consequently, deeper nodes tend to explore increasingly specific,
complementary, or orthogonal semantic facets of the original query,
leading to broader and more structured coverage
of the underlying subtopic space.

\paragraph{Retrieval Agent: Adaptive Retrieval Routing.}
Given a planned query $x_\ell$,
the Retrieval Agent adaptively selects
an appropriate retrieval strategy:
\begin{equation}
r_\ell \sim \pi_{\mathrm{ret}}(x_\ell)
\end{equation}
where $r_\ell$ may correspond to lexical retrieval (e.g., BM25),
dense retrieval, or a hybrid configuration.
Conditioned on the selected strategy,
the agent retrieves a candidate evidence set:
\begin{equation}
\mathcal{D}_\ell
=
\mathrm{Retrieve}(x_\ell, r_\ell, \mathcal{D})
\end{equation}

The selection policy accounts for both
the semantic characteristics of the planned query
(e.g., lexical specificity versus abstraction)
and contextual signals accumulated along the expansion path.
By routing different branches of the retrieval tree
to complementary retrieval signals,
the framework avoids early commitment to a single retriever
and enables robust evidence discovery
across heterogeneous semantic facets.

\paragraph{Augmented Agent: Budget-aware Evidence Normalization.}
Raw retrieval outputs are often noisy, redundant,
and may substantially exceed the available context budget.
The Augmented Agent performs \emph{budget-aware evidence normalization},
transforming heterogeneous retrieval results
into a compact, structured, and information-dense evidence set
suitable for downstream expansion.
This process may involve lightweight re-ranking,
extractive or abstractive compression,
and explicit redundancy reduction,
all subject to a predefined budget constraint.
To ensure robustness,
abstractive compression gracefully falls back
to extractive normalization when confidence is insufficient.
All retained evidence preserves source identifiers,
maintaining full traceability to the original documents.

Formally, given retrieved documents $\mathcal{D}_\ell$
and a context budget $B$,
the normalization operator is defined as
\begin{equation}
\mathcal{E}_\ell
=
\mathcal{G}_{\mathrm{aug}}
\bigl(
q,\,
\mathcal{D}_\ell,\,
B
\bigr),
\quad
\text{s.t. }
\sum_{e \in \mathcal{E}_\ell} \mathrm{len}(e) \le B
\label{eq:augmented_normalization}
\end{equation}
where $\mathcal{E}_\ell$ denotes the normalized evidence set.
The operator jointly optimizes relevance to $q$,
information density,
and redundancy control,
while strictly enforcing budget feasibility.

\paragraph{Child Node Generation.}
Conditioned on the normalized evidence $\mathcal{E}_\ell$,
the system generates multiple child nodes,
each corresponding to a distinct semantic refinement
grounded in retrieved evidence.
Each child node is associated with a concise textual description
and is immediately incorporated into the retrieval tree,
enabling further controlled expansion
and downstream evaluation.

\paragraph{Reflection Agent: Structural Self-correction.}
Recursive expansion without feedback may lead to
semantic drift,
over-specialization,
or structurally unproductive branches.
To mitigate these issues,
we introduce a Reflection Agent that performs
\emph{retrieval-grounded structural self-correction}
during retrieval tree construction.
After a batch of child nodes is expanded,
the agent evaluates each newly induced root-to-node path
$p_{\ell+1}^i = (v_0,\dots,v_\ell,v_{\ell+1}^i)$
by issuing a lightweight retrieval using $p_{\ell+1}^i$ as query context,
and analyzes the resulting evidence distribution $\mathcal{D}_{\ell+1}^i$
in terms of relevance concentration,
semantic coherence,
and contribution to overall coverage.
Based on this retrieval feedback,
the agent selects a localized structural action
$a_{\ell+1}^i \in \{\textsc{Add}(v_{\ell+1}^i),\textsc{Prune}(v_{\ell+1}^i),\textsc{Revise}(v_{\ell+1}^i)\}$
to adjust the tree structure.
The Reflection Agent operates purely at the structural level,
suppressing redundant or low-yield branches
and improving balance and coverage.
Formally, the reflection is defined as:
\begin{equation}
\begin{aligned}
a_{\ell+1}^i
&\sim
\mathrm{Reflect}\bigl(\mathcal{T},\; p_{\ell+1}^i,\; \mathcal{D}_{\ell+1}^i\bigr) \\
\mathcal{T}
&\leftarrow
\Phi\bigl(\mathcal{T},\; v_{\ell+1}^i,\; a_{\ell+1}^i\bigr)
\end{aligned}
\label{eq:reflection_update}
\end{equation}

where $\mathrm{Reflect}(\cdot)$ denotes a retrieval-grounded structural policy
and $\Phi(\cdot)$ applies the corresponding localized transformation.

\paragraph{Discussion.}
Stage~I formulates retrieval tree construction
as a controlled, role-specialized agentic process,
enabling systematic semantic exploration
while explicitly managing redundancy,
semantic drift,
and structural imbalance.
The resulting retrieval tree is constructed once per query topic
and serves as a persistent, reusable representation
of the query’s semantic space,
allowing Stage~II to adapt retrieval behavior
to evolving news distributions
without re-invoking expensive agentic reasoning.

\subsection{Dynamic Adaptive Subtree Selection}
\label{subsec:stage2}

Stage~I constructs a  retrieval tree
that captures diverse semantic interpretations of a query.
However, indiscriminately applying the full tree
is neither necessary nor optimal
under evolving news distributions.
As surface realizations, topical emphasis,
and relevance patterns shift over time,
different substructures of the tree
may exhibit substantially different retrieval effectiveness.
To address this mismatch,
we introduce a \emph{dynamic adaptive subtree selection} mechanism
that adapts retrieval behavior
by selecting an appropriate subtree,
without re-running expensive semantic exploration.

As formalized in Section~\ref{subsec:problem},
the ideal objective is to maximize
recall-oriented news retrieval performance
over the full corpus.
Directly optimizing Eq.~(\ref{eq:objective}) is not only computationally infeasible,
but also misaligned with the structural nature of the problem:
the goal of Stage~II is not to estimate absolute recall,
but to reliably \emph{rank} alternative semantic structures
under stable and bounded cost. Instead, Stage~II focuses on
\emph{efficiently approximating} this objective
by enabling reliable \emph{relative comparison}
among candidate subtrees
under stable and bounded computational cost.

\paragraph{Evaluation Proxy Construction.}
We first instantiate the retrieval function
$\mathcal{R}(q,\mathcal{T})$
using the full retrieval tree
to obtain a global ranking.
Evaluation is then restricted to
a compact candidate pool
\begin{equation}
\mathcal{C}
=
\mathrm{Top}_{M}\bigl(\mathcal{R}(q,\mathcal{T})\bigr)
\end{equation}
which concentrates documents
most strongly supported by the current semantic structure
while keeping downstream computation tractable.
Each document in $\mathcal{C}$
is represented by a dense embedding,
enabling efficient similarity-based operations.

To avoid overemphasizing dominant semantic modes,
we further construct a diversity-aware evaluation proxy.
Specifically, we perform coarse-grained clustering
over document embeddings in $\mathcal{C}$
and allocate evaluation quotas proportionally across clusters.
Let $\{\mathcal{C}_1, \dots, \mathcal{C}_K\}$
denote the resulting partition.
Given a target proxy size $N_{\mathrm{eval}}$,
cluster-level quotas are defined as
\begin{equation}
n_k
=
\left\lfloor
N_{\mathrm{eval}}
\cdot
\frac{|\mathcal{C}_k|}{|\mathcal{C}|}
\right\rfloor,
\qquad k = 1,\dots,K
\label{eq:cluster_quota}
\end{equation}
Within each cluster $\mathcal{C}_k$,
documents are selected by global rank,
yielding $\mathcal{P}_k = \mathrm{Top}_{n_k}(\mathcal{C}_k)$.
The final evaluation proxy
$\mathcal{P}_{\mathrm{eval}}$
is initialized by aggregating
$\{\mathcal{P}_1, \dots, \mathcal{P}_K\}$.
The remaining slots, if any, are filled according to the global ranking
until the proxy reaches the target size $N_{\mathrm{eval}}$.
When clustering quality degrades,
the procedure gracefully falls back
to stratified rank-based selection.

Subtree selection is intentionally evaluated on a compact proxy rather than the full corpus,
yet this approximation remains well aligned
with the ideal objective.
The candidate pool is derived
from full-tree retrieval,
which already emphasizes high-coverage regions,
while diversity-aware sampling
preserves representation across multiple semantic facets.
Crucially, subtree selection is inherently \emph{relative}:
as long as proxy-based estimates
preserve the ordering among candidate subtrees,
the selected structure remains stable,
even if absolute recall values are imperfect.

\paragraph{Subtree Evaluation and Selection.}
As formulated in Section~\ref{subsec:problem},
the ideal objective is to select a subtree
that maximizes Recall
with respect to the true relevance set $\mathcal{D}^{+}(q)$,
which is infeasible to obtain at scale.
We therefore approximate this objective
using a compact yet semantically diverse evaluation proxy
$\mathcal{P}_{\mathrm{eval}}$,
where documents are weakly annotated by an LLM
as relevant or non-relevant to $q$,
yielding a weakly labeled relevant subset
$\mathcal{P}_{\mathrm{eval}}^{+}(q)$, where
$|\mathcal{P}_{\mathrm{eval}}^{+}(q)| = \alpha$.
For each candidate subtree $\mathcal{T}' \in \mathbb{T}(\mathcal{T})$,
we instantiate the induced path-ensemble retrieval function
$\mathcal{R}(q,\mathcal{T}')$
and estimate its Recall@$\alpha$
on $\mathcal{P}_{\mathrm{eval}}$.
The selected subtree is thus obtained by
\begin{equation}
\mathcal{T}^\star
=
\arg\max_{\mathcal{T}' \in \mathbb{T}(\mathcal{T})}
\mathrm{Recall@}\alpha\bigl(
\mathcal{R}(q, \mathcal{T}')|_{\mathcal{P}_{\mathrm{eval}}},
\mathcal{P}_{\mathrm{eval}}^+(q)
\bigr)
\end{equation}

which serves as a practical approximation
to the ideal objective in Eq.~(\ref{eq:objective}).

\paragraph{Discussion.}
Dynamic adaptive subtree selection
enables rapid adaptation
to evolving news distributions
by reconfiguring semantic structure
instead of regenerating semantic expansions.
By selecting reusable subtrees,
the framework amortizes expensive exploration in Stage~I
and achieves efficient, robust adaptation in Stage~II,
maintaining high coverage,
low latency,
and strong interpretability
for large-scale news retrieval.
Stage~II uses LLM-based weak labels only for relative subtree comparison,
so label noise mainly affects selection stability rather than directly biasing retrieval.
To reduce proxy bias, we build the proxy via embedding-based clustering
and quota-based sampling, encouraging coverage of diverse semantic modes
beyond dominant full-tree retrieval directions.
This design makes agentic retrieval viable in production,
where semantic exploration is costly,
news distributions evolve continuously,
and retrieval policies must adapt under strict latency constraints.

\section{Experiments}
\label{sec:experiments}

\begin{table*}[t]
\centering
\small
\setlength{\tabcolsep}{3pt}
\caption{Performance on the Syft news dataset and BEIR benchmark datasets.
For each method, the first row shows Recall@100 and the second row shows NDCG@10.
Best results are highlighted in \textbf{bold}, second-best are \underline{underlined}.}
\label{tab:combined_results}

\begin{tabular}{c|ccccccc|c|ccccc|c|c}
\toprule
\multirow{2}{*}{\textbf{Method}} &
\multicolumn{7}{c|}{\textbf{Syft News}} &
\multirow{2}{*}{\textbf{Avg}} &
\multicolumn{5}{c|}{\textbf{BEIR}} &
\multirow{2}{*}{\textbf{Avg}} &
\multirow{2}{*}{\textbf{Total Avg}} \\
\cmidrule(lr){2-8} \cmidrule(lr){10-14}
 & 6.11 & 6.12 & 6.13 & 6.14 & 6.15 & 6.29 & 6.30
 &  &
 Trec-Covid & NfCorpus & FiQA & SciFact & SciDocs
 &  &  \\
\midrule

\multirow{2}{*}{ReAct} & 0.409 & 0.392 & 0.403 & 0.399 & \underline{0.345} & 0.295 & 0.375 & 0.374
 & 0.638 & 0.422 & 0.817 & \underline{0.983} & 0.495 & 0.671 & 0.498 \\[-2pt]
 & 0.562 & 0.599 & \underline{0.727} & 0.716 & \underline{0.764} & 0.794 & 0.655 & 0.688
 & 0.753 & \underline{0.439} & 0.454 & 0.775 & 0.228 & 0.530 & 0.622 \\
\cmidrule(lr){1-16}
\multirow{2}{*}{Reflexion} & \underline{0.426} & 0.395 & 0.405 & 0.372 & 0.320 & 0.282 & 0.344 & 0.364
 & 0.631 & 0.406 & 0.808 & \textbf{0.990} & 0.478 & 0.663 & 0.488 \\[-2pt]
 & 0.561 & 0.611 & 0.726 & 0.713 & 0.720 & 0.768 & 0.647 & 0.678
 & 0.758 & 0.427 & 0.441 & 0.756 & 0.209 & 0.518 & 0.611 \\
\cmidrule(lr){1-16}
\multirow{2}{*}{FreshLLMs} & 0.374 & 0.354 & 0.323 & 0.313 & 0.274 & 0.253 & 0.319 & 0.316
 & 0.593 & 0.404 & 0.800 & 0.962 & 0.474 & 0.647 & 0.454 \\[-2pt]
 & 0.450 & 0.512 & 0.578 & 0.581 & 0.599 & 0.679 & 0.527 & 0.561
 & 0.732 & 0.393 & 0.402 & 0.710 & 0.202 & 0.488 & 0.530 \\
\cmidrule(lr){1-16}
\multirow{2}{*}{Self-RAG} & 0.382 & 0.408 & 0.373 & 0.362 & 0.322 & 0.281 & 0.349 & 0.354
 & 0.629 & 0.425 & 0.818 & 0.957 & 0.496 & 0.665 & 0.484 \\[-2pt]
 & 0.545 & 0.618 & 0.666 & 0.685 & 0.708 & 0.788 & 0.613 & 0.660
 & 0.784 & 0.436 & 0.455 & 0.717 & 0.225 & 0.523 & 0.603 \\
\cmidrule(lr){1-16}
\multirow{2}{*}{RAPTOR} & 0.357 & 0.410 & 0.358 & 0.359 & 0.293 & 0.257 & 0.327 & 0.337
 & 0.566 & 0.410 & 0.789 & 0.977 & 0.490 & 0.646 & 0.466 \\[-2pt]
 & 0.519 & 0.591 & 0.654 & 0.662 & 0.676 & 0.730 & 0.612 & 0.635
 & 0.737 & 0.409 & 0.420 & \textbf{0.785} & 0.212 & 0.513 & 0.584 \\
\cmidrule(lr){1-16}
\multirow{2}{*}{HyDE} & 0.382 & 0.421 & 0.383 & 0.358 & 0.296 & 0.264 & 0.343 & 0.350
 & \underline{0.665} & 0.414 & \underline{0.838} & 0.977 & \underline{0.516} & \underline{0.682} & 0.488 \\[-2pt]
 & 0.538 & 0.593 & 0.676 & 0.646 & 0.661 & 0.730 & 0.610 & 0.636
 & \textbf{0.858} & 0.428 & \underline{0.501} & \underline{0.778} & \underline{0.242} & \underline{0.562} & 0.605 \\
\cmidrule(lr){1-16}
\multirow{2}{*}{KG$^2$RAG} & 0.413 & 0.414 & \underline{0.411} & \underline{0.404} & 0.324 & \underline{0.302} & \underline{0.391} & \underline{0.380}
 & 0.636 & \textbf{0.431} & 0.825 & \underline{0.983} & 0.498 & 0.675 & \underline{0.503} \\[-2pt]
 & \underline{0.583} & 0.621 & 0.714 & \underline{0.730} & 0.750 & \underline{0.795} & \underline{0.674} & \underline{0.695}
 & 0.747 & \textbf{0.440} & 0.447 & 0.747 & 0.223 & 0.521 & \underline{0.623} \\
\cmidrule(lr){1-16}
\multirow{2}{*}{GraphRAG} & 0.421 & \underline{0.427} & 0.379 & 0.367 & 0.314 & 0.278 & 0.358 & 0.363
 & 0.536 & 0.381 & 0.663 & 0.942 & 0.458 & 0.596 & 0.460 \\[-2pt]
 & 0.548 & \underline{0.632} & 0.677 & 0.662 & 0.715 & 0.728 & 0.616 & 0.654
 & 0.589 & 0.365 & 0.272 & 0.640 & 0.189 & 0.411 & 0.553 \\
\cmidrule(lr){1-16}
\cellcolor{gray!20} & \cellcolor{gray!20}\textbf{0.555} & \cellcolor{gray!20}\textbf{0.579} & \cellcolor{gray!20}\textbf{0.507} & \cellcolor{gray!20}\textbf{0.473} & \cellcolor{gray!20}\textbf{0.408} & \cellcolor{gray!20}\textbf{0.360} & \cellcolor{gray!20}\textbf{0.446} & \cellcolor{gray!20}\textbf{0.475}
 & \cellcolor{gray!20}\textbf{0.667} & \cellcolor{gray!20}\underline{0.426} & \cellcolor{gray!20}\textbf{0.852} & \cellcolor{gray!20}0.977 & \cellcolor{gray!20}\textbf{0.521} & \cellcolor{gray!20}\textbf{0.689} & \cellcolor{gray!20}\textbf{0.564} \\[-2pt]
\multirow{-2}{*}{\cellcolor{gray!20}\textbf{DynaTree}} & \cellcolor{gray!20}\textbf{0.664} & \cellcolor{gray!20}\textbf{0.709} & \cellcolor{gray!20}\textbf{0.796} & \cellcolor{gray!20}\textbf{0.786} & \cellcolor{gray!20}\textbf{0.791} & \cellcolor{gray!20}\textbf{0.844} & \cellcolor{gray!20}\textbf{0.709} & \cellcolor{gray!20}\textbf{0.757}
 & \cellcolor{gray!20}\underline{0.855} & \cellcolor{gray!20}\textbf{0.440} & \cellcolor{gray!20}\textbf{0.548} & \cellcolor{gray!20}\textbf{0.785} & \cellcolor{gray!20}\textbf{0.244} & \cellcolor{gray!20}\textbf{0.575} & \cellcolor{gray!20}\textbf{0.681} \\
\bottomrule
\end{tabular}
\end{table*}

In this section, we evaluate the proposed agentic semantic expansion framework
under realistic industrial constraints,
using both \emph{offline controlled experiments}
and \emph{online platform evaluations}.
Our evaluation focuses on assessing
the robustness and effectiveness of structured agentic retrieval
across complementary retrieval metrics and deployment settings.
Our experimental design is structured around the following research questions:
\begin{itemize}
    \item \textbf{RQ1:} Does the agentic retrieval tree
    consistently improve retrieval effectiveness
    over strong RAG and agentic baselines
    across both Syft news data and public benchmarks?

    \item \textbf{RQ2:} How do different design choices
    within the agentic framework contribute to overall effectiveness,
    including individual agentic components
    and best-subtree selection
    compared to using the full retrieval tree?

    \item \textbf{RQ3:} How sensitive is the proposed framework to key design parameters,
such as retrieval tree depth and size of the evaluation proxy
used for dynamic subtree selection?

    \item \textbf{RQ4:} Does the proposed method
    lead to measurable improvements
    in real-world user experience
    on our online news retrieval platform?
\end{itemize}

\subsection{Experimental Setup}
\label{subsec:exp_setup}

\paragraph{\textbf{Datasets.}}
Our primary evaluation corpus is a proprietary news dataset collected by our project team (\textbf{Syft}),
covering news stories produced on June 11--15, June 29, and June 30, 2025
across diverse real-world topics (See Appendix~\ref{sec:appendix_syft_data} for details).
We define 50 query themes for offline evaluation, for which relevance judgments are
annotated against the news corpus of each corresponding day, and for online evaluation, we further introduce 30 complex themes
and 20 long-tail themes to reflect more challenging real-world queries.
To assess generalization beyond the internal domain,
we also evaluate our approach on 5 publicly available datasets
from the BEIR benchmark~\cite{thakur2021beirheterogenousbenchmarkzeroshot},
which provides standardized retrieval tasks across diverse domains.

\paragraph{\textbf{Evaluation Metrics.}}
Offline retrieval performance is evaluated following BEIR protocol ~\cite{thakur2021beirheterogenousbenchmarkzeroshot},
using \textbf{Recall@100} and \textbf{NDCG@10}.
For datasets with more than 100 relevant documents per query
(e.g., Syft news and Trec-Covid in BEIR),
we report \textbf{capped Recall@100} as defined in BEIR.
Recall@100 measures high-recall retrieval effectiveness,
while NDCG@10 captures ranking quality with position bias~\cite{pmlr-v30-Wang13}.
Online performance is evaluated using a platform-provided
\textbf{survival rate},
which reflects whether retrieved items are retained
after downstream filtering and quality control,
a standard proxy for content utility in industrial retrieval systems.

\paragraph{\textbf{Baselines.}}
We compare our method against a diverse and representative set of strong baselines
spanning multiple paradigms of LLM-based retrieval and semantic expansion.
For offline evaluation, we include
\emph{agentic and reasoning-based} methods
(\textbf{ReAct}~\cite{yao2023reactsynergizingreasoningacting},
\textbf{Reflexion}~\cite{10.5555/3666122.3666499},
\textbf{FreshLLMs}~\cite{vu-etal-2024-freshllms},
and \textbf{Self-RAG}~\cite{ICLR2024_25f7be96}),
a representative \emph{LLM-based query expansion} approach
(\textbf{HyDE}~\cite{gao-etal-2023-precise}),
as well as \emph{structure- and knowledge-aware retrieval} methods,
including \textbf{RAPTOR}~\cite{DBLP:conf/iclr/SarthiATKGM24},
\textbf{GraphRAG}~\cite{edge2025localglobalgraphrag},
and the recent \textbf{KG$^2$RAG}~\cite{zhu-etal-2025-knowledge} framework,
which incorporates explicit knowledge graph structures
into RAG.  
To better relate DynaTree to classical stream-based retrieval, we evaluate KBA-style lexical and nugget-based baselines on the Syft news benchmark, including BM25, expanded BM25, nugget-based BM25, and nugget coverage. These baselines are intended as diagnostic comparisons for topic-level news retrieval.
For online evaluation, we further compare against five
production-grade recallers deployed within the same product line,
including \textit{relation-based recall},
\textit{decompositional recall},
\textit{direct semantic recall},
\textit{collaborative proxy recall},
and \textit{web search recall}. (See Appendix~\ref{sec:appendix_recallers} for details).

\paragraph{\textbf{Other Settings.}}
Unless otherwise specified, all agentic components use
\textbf{Gemini-2.5-Flash}~\cite{comanici2025gemini}.
Documents and semantic expansions are encoded with
\textbf{text-embedding-3-large},
and cosine similarity is used for matching.
All experiments enforce a fixed semantic expansion budget,
including constraints on retrieval tree depth and total node count.
To isolate the effect of dynamic adaptive subtree selection,
the retrieval tree is constructed once and reused across experiments.
For the Syft news benchmark,
the tree is built on the June~29,~2025 snapshot
and reused for all subsequent dates;
Stage~II performs subtree selection
using a fixed proxy set of size
$N_{\mathrm{eval}} = 200$
under a unified evaluation protocol,
with a three-level retrieval tree.
In contrast, BEIR benchmarks do not involve subtree selection:
Stage~II is designed to adapt retrieval behavior under evolving news distributions,
whereas BEIR provides a static offline corpus.
We therefore directly use the full retrieval tree,
constructed on a restricted corpus of the top 10{,}000
query-relevant documents selected via dense retrieval,
to ensure a stable and reproducible evaluation setting.
All BEIR datasets also use a three-level retrieval tree,
except SciDocs and FiQA,
which adopt a two-level tree due to their larger query sets.
We report absolute construction cost and scaling analysis in
Appendix~\ref{app:cost_analysis}.

\subsection{Offline Evaluation}
\label{subsec:offline}

\subsubsection{\textbf{Main Results (RQ1)}}

Table~\ref{tab:combined_results} reports the main offline results
on the Syft news benchmark and the BEIR datasets.
On the Syft news corpus,
our method consistently achieves the strongest performance
in both Recall@100 and NDCG@10 across all seven evaluation dates,
demonstrating superior early-stage ranking quality
and robust recall under realistic retrieval cutoffs.
Notably, the improvements remain stable
across temporally distinct news snapshots,
indicating that the reusable agentic retrieval tree
effectively adapts to evolving relevance distributions
without repeated semantic exploration. On the BEIR benchmarks,
our approach also attains the best overall performance on average,
outperforming strong query-expansion and agentic RAG baselines
across heterogeneous domains and query types.
These results suggest that the proposed retrieval tree representation generalizes well beyond the news domain without task-specific tuning, while Stage-II subtree selection further improves adaptation in dynamic news settings.

\begin{table}[htbp]
\centering
\small
\caption{Comparison with KBA-style lexical and nugget-based baselines on the Syft news benchmark. Results are averaged over seven daily snapshots.}
\label{tab:kba_style}
\begin{tabular}{lcc}
\toprule
\textbf{Method} & \textbf{Recall@100} & \textbf{NDCG@10} \\
\midrule
BM25 & 0.144 & 0.353 \\
BM25(expanded) & \underline{0.158} & \underline{0.389} \\
Nugget-BM25 & 0.148 & 0.373 \\
Nugget coverage & 0.112 & 0.293 \\
\midrule
\textbf{DynaTree} & \textbf{0.475} & \textbf{0.757} \\
\bottomrule
\end{tabular}
\end{table}

\paragraph{Alignment with KBA-style Stream Retrieval.}
DynaTree is not designed for the original KBA objective of entity-centric delta detection, but it shares with KBA-style evaluations the need to retrieve relevant information from continuously evolving document streams. The key difference is that DynaTree targets topic-level, latency-constrained news retrieval for downstream ranking rather than knowledge-base updates. To examine this connection, we introduce KBA-style lexical and nugget-based baselines on the Syft news benchmark, including BM25, expanded BM25, nugget-BM25, and direct nugget coverage. As shown in Table~\ref{tab:kba_style}, DynaTree achieves 0.475 Recall@100 and 0.757 NDCG@10 on the seven-day average, substantially outperforming the strongest KBA-style baseline, which achieves 0.158 Recall@100 and 0.389 NDCG@10. These results suggest that flat lexical expansion or isolated nugget matching is insufficient for topic-level semantic coverage in dynamic news streams, while DynaTree benefits from structured semantic expansion and adaptive subtree selection.

\subsubsection{\textbf{Ablation Study (RQ2)}}

\paragraph{\textbf{Impact of Agentic Components.}}
\label{exp:ablation1}
Table~\ref{tab:ablation} reports both single-module ablations and Shapley-based attributions, with all results averaged over seven evaluation days on the Syft news benchmark using the full tree.
Single ablation reflects the effect of removing a module only in the full configuration,
while Shapley values measure average marginal contributions over all component subsets
and thus capture interaction effects.
Empirically, removing any module degrades overall performance
(e.g., w/o P lowers Recall@100 from 37.50 to 36.82,
and w/o F lowers NDCG@10 from 67.82 to 67.05),
confirming that all agents are beneficial.
Shapley analysis shows that Planning contributes most to Recall (+0.253),
whereas Reflection contributes most to NDCG (+0.566),
while Retrieval and Augment exhibit smaller or mixed effects,
consistent with strong interdependencies.
Detailed attribution protocol and results of $2^4$ configs are provided in
Appendix~\ref{sec:appendix_shapley}.

\begin{table}[htbp]
\centering
\small
\setlength{\tabcolsep}{4pt}
\caption{Ablation study and Shapley value analysis.
All reported values are in percentage (\%).
Left: performance degradation when removing individual modules.
Right: marginal contributions estimated by Shapley values.}

\label{tab:ablation}
\begin{tabular}{ccc|ccc}
\toprule
\multicolumn{3}{c|}{\textbf{Ablation Study}} &
\multicolumn{3}{c}{\textbf{Shapley Contribution}} \\
\cmidrule(lr){1-3} \cmidrule(lr){4-6}
\textbf{Config} & \textbf{R@100} & \textbf{N@10} &
\textbf{Module} & \textbf{R@100} & \textbf{N@10} \\
\midrule
w/o P     & 36.82 & 67.19 & P & \textbf{+0.253} & +0.146 \\
w/o R     & 36.62 & 67.48 & R & +0.156 & $-$0.019 \\
w/o A     & 37.12 & 67.93 & A & +0.008 & +0.038  \\
w/o F     & 36.73 & 67.05 & F & +0.094 & \textbf{+0.566}      \\
Full      & 37.50 & 67.82 & Sum &  +0.510     & +0.730\\
\bottomrule
\end{tabular}
\end{table}

\begin{figure}[htbp]
    \centering
    \setlength{\abovecaptionskip}{4pt}
    \setlength{\belowcaptionskip}{0pt}
    \includegraphics[width=\linewidth]{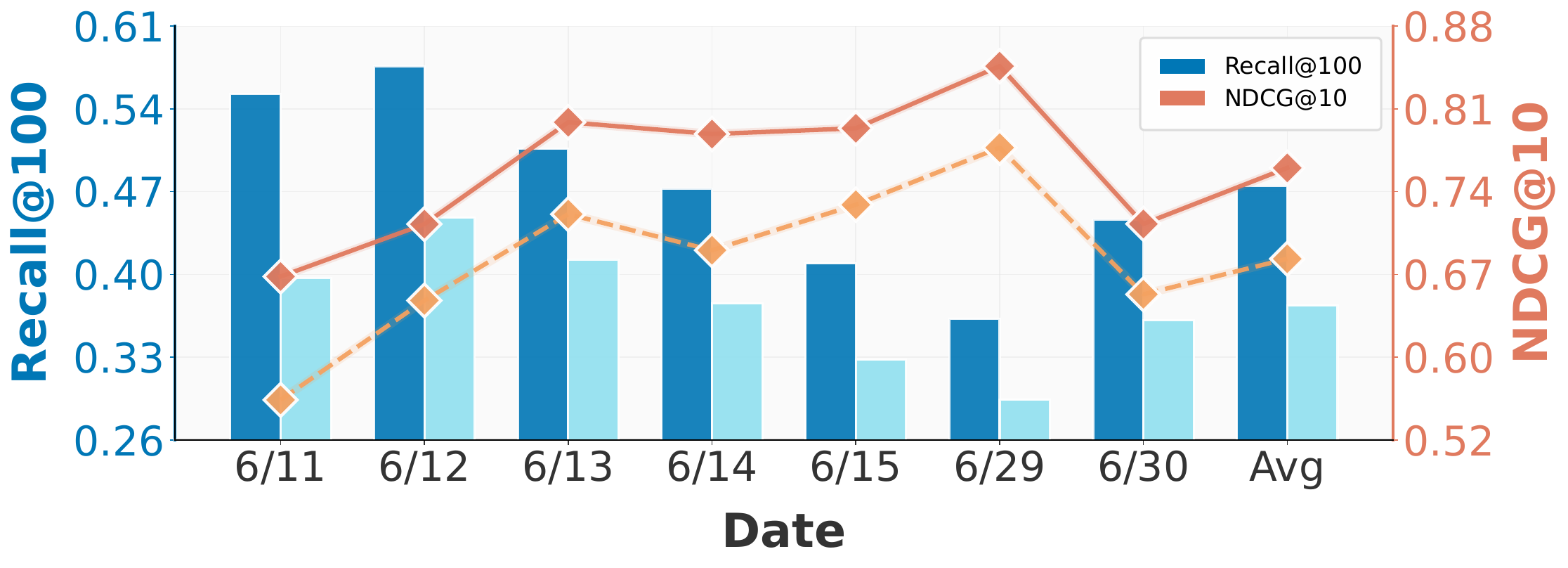}
    \caption{
    Effect of subtree selection on Syft news.
    Dark colors indicate full setting, while light colors indicate ablation.
}

    \label{fig:ablation_stage2}
\end{figure}

\paragraph{\textbf{Effect of Stage-II Subtree Selection.}}
Figure~\ref{fig:ablation_stage2} compares retrieval performance
between using the full retrieval tree
and the dynamically selected best subtree
across seven days of news data.
Stage-II subtree selection consistently improves
both Recall@100 and NDCG@10 on all evaluated dates.
The gains are stable over time,
indicating that selecting a compact, adaptive subtree
is more effective than indiscriminately using the full tree.
These results demonstrate that Stage-II
successfully adapts retrieval behavior
to temporal variations in news distributions,
achieving higher recall and better ranking quality.

\paragraph{\textbf{Uniform vs. Weighted Aggregation.}}
Across seven days of news data, we compare two retrieval uses of the dynamically selected best subtree:
uniform averaging versus softmax weighting of article--path similarities.
Uniform averaging encourages balanced coverage across selected semantic paths,
while Stage~II already provides implicit structure-level weighting through subtree selection.
Uniform aggregation slightly outperforms softmax weighting over seven daily snapshots
(0.475/0.757 vs. 0.464/0.752 in Recall@100/NDCG@10).
Full results are reported in Appendix~\ref{app:uniform_weighted}.

\subsubsection{\textbf{Sensitivity Analysis (RQ3)}}
\begin{figure}[htbp]
    \centering
    \includegraphics[width=\linewidth]{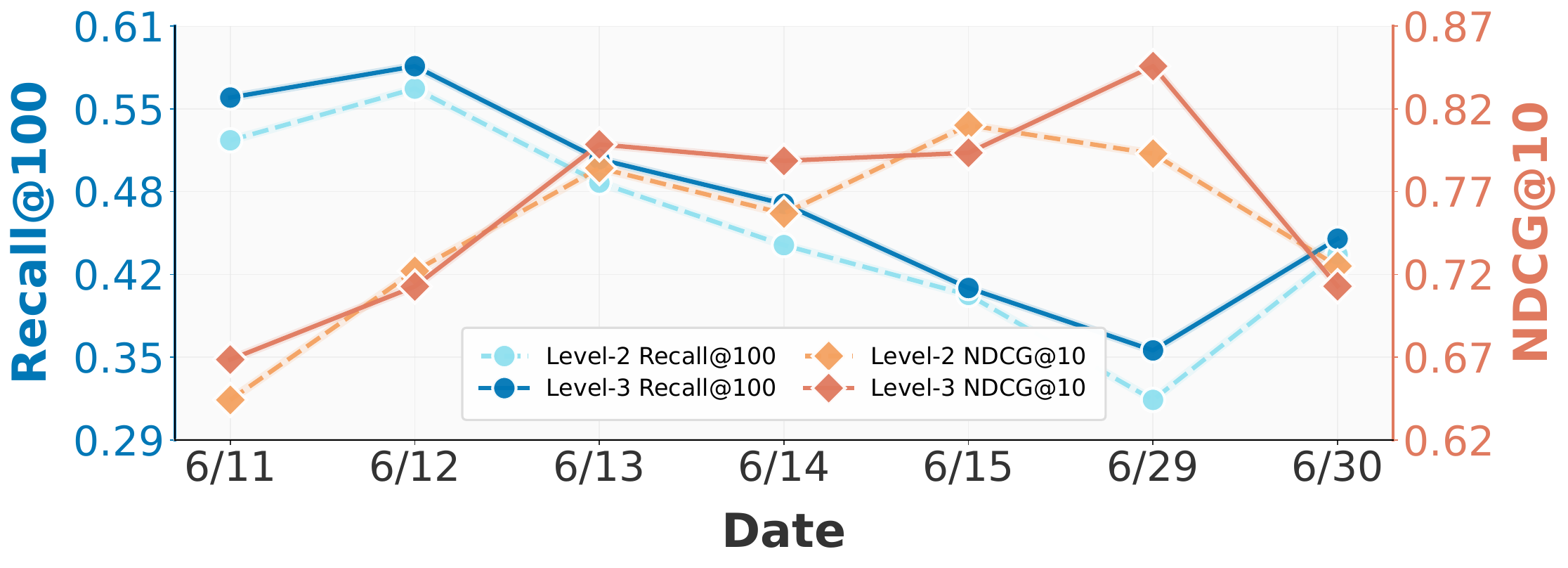}
    \caption{
    Recall@100 and NDCG@10 comparison between Level-2 and Level-3 retrieval trees over seven days.
    }
    \label{fig:recall_level_comparison}
\end{figure}

\paragraph{\textbf{Effect of Retrieval Tree Depth.}}
Figure~\ref{fig:recall_level_comparison} examines the effect of retrieval tree depth
by comparing Level-2 and Level-3 best subtrees across seven evaluation days on Syft news.
Across all days, Level-3 consistently achieves higher Recall@100 and NDCG@10 than Level-2,
with more stable gains under distributional shifts.
This suggests that deeper retrieval trees provide a richer semantic structure,
which becomes effective when adaptive subtree selection isolates
high-utility branches while suppressing redundant or noisy expansions.

\paragraph{\textbf{Evaluation Proxy Size Analysis.}}
\label{para:pilot_size}

\begin{figure}[htbp]
    \centering
    \includegraphics[width=\linewidth]{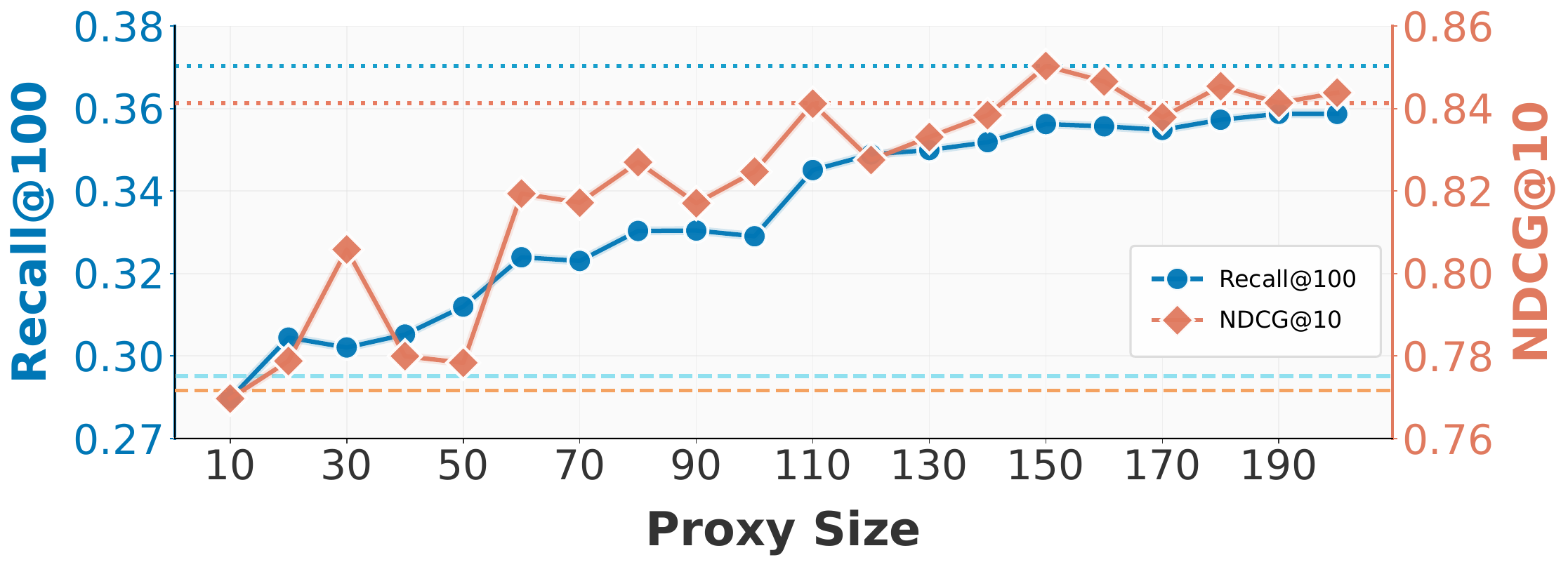}
    \caption{Effect of evaluation proxy size on Stage-II performance. The dashed lines indicate the no–Stage-II lower bound and the full-proxy upper bound.}
    \label{fig:pilot_size}
\end{figure}

We analyze how the size of the evaluation proxy
used for subtree selection
affects Stage~II performance on the June~29 news snapshot.
As the proxy size increases,
both Recall@100 and NDCG@10 improve steadily,
indicating more reliable relative comparisons
among candidate subtrees.
Very small proxies yield unstable estimates,
while performance gains diminish beyond approximately 150 samples.
With 200 proxy examples,
Stage~II achieves Recall@100 of 0.360
and NDCG@10 of 0.844,
closely approaching the upper-bound performance (using full data as proxy).
We also observe strong agreement between LLM-based weak labels
and human annotations on sampled proxy examples,
suggesting that the proxy provides a reliable signal
for relative subtree comparison. 
Overall, these results demonstrate that Stage~II is data-efficient
and remains robust under realistic news distributions,
even when guided by a compact evaluation proxy
(Figure~\ref{fig:pilot_size}).

\subsection{Online Deployment and Evaluation (RQ4)}
\label{subsec:online}

\paragraph{\textbf{Online Setting.}}
We evaluate DynaTree in \textbf{Syft}, a production-level AI-powered news platform developed by Orion Arm AI that delivers personalized, topic-centric news channels to a global user base.
Syft aggregates news from diverse international sources, deduplicates overlapping reports, and produces concise summaries, organizing content around persistent user-defined interest channels (e.g., technology policy, financial markets, or regional affairs) rather than a generic feed.
From a retrieval perspective, Syft operates a large-scale online pipeline where multiple production-grade recallers jointly generate candidates, including relation-based recall (RelRec), decompositional recall (Decomp), direct semantic recall (SemRec), collaborative proxy recall (Proxy), and web search recall (WSRec).
Candidates from all recallers are merged and processed by a shared downstream filtering and ranking pipeline under identical online constraints.
The channel-centric and continuously evolving nature of Syft induces highly diverse and dynamic retrieval demands, making it a realistic and high-stakes environment for evaluating online news retrieval quality, where recall effectiveness directly impacts user experience.

\paragraph{\textbf{A/B Testing.}}
To assess the effectiveness of online daily subtree adaptation,
we conduct an A/B test in the deployed DynaTree system.
The full agentic retrieval tree is constructed once
using the January~27,~2026 news corpus,
covering 100 representative topic channels,
and is kept fixed throughout the experiment,
ensuring that both variants share an identical semantic structure
and differ only in subtree selection.
User traffic for each channel is evenly split:
Port~A uses a \emph{static} strategy,
where query expansions are generated from a best-performing subtree
identified offline and kept unchanged,
while Port~B applies a \emph{dynamic} strategy,
selecting and updating subtrees daily
based on real-time news distribution and retrieval feedback.
Both variants are evaluated alongside existing production-grade recallers
under identical online constraints,
sharing the same downstream filtering and ranking pipeline,
and are measured using the platform \emph{survival rate},
defined as the fraction of retrieved news
that passes downstream quality control.

\paragraph{\textbf{Results and Analysis.}}
Table~\ref{tab:online_ab_results} reports online A/B testing results
from Jan.~28 to Feb.~6,~2026 measured by survival rate.
Across all days, Port~B, which performs daily online subtree selection,
consistently outperforms Port~A with a fixed offline-selected subtree,
indicating that subtree effectiveness varies over time
and static configurations are insufficient
under evolving news distributions.
Port~B also achieves the best performance among all production recallers
on every reported day, often with a clear margin,
demonstrating the benefit of dynamic, structure-aware adaptation
over heterogeneous but static recall strategies.

Beyond controlled A/B evaluation,
DynaTree has been fully deployed in real news topic channels
and serves live user traffic.
The same retrieval pipeline directly delivers content
to subscribed users,
so the observed online gains reflect
not only system-level improvements
but also tangible user-facing benefits
in content coverage, freshness, and relevance.
In practice, user interests are organized into a set of persistent topic channels,
where each tree is constructed once and reused online,
so offline cost scales with maintained channels rather than raw query volume.
For unseen or cold-start topics,
the system falls back to existing low-latency production recallers
while queuing offline tree construction for future reuse.
These results provide strong evidence
of the practical effectiveness and deployability
of DynaTree in real-world news retrieval systems.

\begin{table}[t]
\centering
\small
\setlength{\tabcolsep}{4pt}
\caption{Online A/B testing results,
measured by survival rate.}
\label{tab:online_ab_results}
\begin{tabular}{c|cc|ccccc}
\toprule
\textbf{Date} &
\multicolumn{2}{c|}{\textbf{DynaTree}} &
\multicolumn{5}{c}{\textbf{Prod. Recallers}} \\
\cmidrule(lr){2-3} \cmidrule(lr){4-8}
 & Port A & Port B & RelRec & Decomp & SemRec & Proxy & WSRec \\
\midrule
1.28 & 0.39 & \textbf{0.72} & 0.21 & 0.21 & 0.01 & 0.10 & 0.60 \\
1.29 & 0.48 & \textbf{0.68} & 0.21 & 0.12 & 0.01 & 0.10 & 0.33 \\
1.30 & 0.46 & \textbf{0.70} & 0.22 & 0.12 & 0.01 & 0.11 & / \\
1.31 & 0.48 & \textbf{0.71} & 0.25 & 0.11 & 0.01 & 0.07 & 0.00 \\
2.01 & 0.32 & \textbf{0.73} & 0.17 & 0.12 & 0.01 & 0.08 & 0.40 \\
2.02 & 0.38 & \textbf{0.59} & 0.16 & 0.14 & 0.01 & 0.09 & 0.25 \\
2.03 & 0.48 & \textbf{0.60} & 0.16 & 0.10 & 0.01 & 0.05 & 0.40 \\
2.04 & 0.49 & \textbf{0.59} & 0.22 & 0.07 & 0.01 & 0.07 & 0.33 \\
2.05 & 0.53 & \textbf{0.69} & 0.26 & 0.11 & 0.01 & 0.04 & 0.00 \\
2.06 & 0.46 & \textbf{0.68} & 0.21 & 0.11 & 0.01 & 0.10 & 0.50 \\
\midrule
\textbf{Avg} & 0.45 & \textbf{0.67} & 0.21 & 0.12 & 0.01 & 0.08 & 0.31 \\
\bottomrule
\end{tabular}
\end{table}

\section{Conclusion}
\label{sec:conclusion}

We propose \textbf{DynaTree}, a two-stage agentic retrieval framework
for effective and efficient news retrieval
under evolving temporal distributions.
By materializing semantic expansion as a reusable retrieval tree
and decoupling its construction from online optimization,
DynaTree amortizes agentic reasoning cost
while enabling lightweight daily adaptation via dynamic subtree selection, guided by a compact evaluation proxy.
Extensive offline experiments and direct online deployment
in the Syft production system
demonstrate consistent improvements over strong RAG and agentic baselines,
showing that persistent semantic structures
can maintain robust recall and ranking quality over time.
These results highlight the practical value of structure-aware,
agentic retrieval for large-scale, real-world news systems
and its potential to deliver more reliable retrieval experiences to users.

\begin{acks}
The SJTU team is partially supported by the National Natural Science Foundation of China
(No. 62502310, No. 62322603, and No. 624B2096).
\end{acks}

\bibliographystyle{ACM-Reference-Format}
\balance
\bibliography{sample-base}


\appendix

\section{Algorithmic Description}
\label{appendix:algorithm}

Algorithm~\ref{alg:dart_rag} presents the complete algorithmic description
of the proposed DynaTree framework,
including agentic retrieval tree construction
and dynamic adaptive subtree selection.

\begin{algorithm}[htbp]
\caption{DynaTree: Agentic Retrieval Tree Construction and Dynamic Subtree Selection}
\label{alg:dart_rag}
\small
\DontPrintSemicolon

\KwIn{
Query $q$; corpus $\mathcal{D}$; max tree depth $L$; 
context budget $B$; proxy size $N_{\mathrm{eval}}$
}
\KwOut{Selected subtree $\mathcal{T}^\star$}

\vspace{0.5em}
\textbf{Stage I: Agentic Retrieval Tree Construction}

Initialize retrieval tree $\mathcal{T}$ with root $v_0 = q$\;
Initialize frontier $\mathcal{F} \leftarrow \{(v_0, p_0)\}$, where $p_0=(v_0)$\;

\For{$\ell \leftarrow 0$ \KwTo $L-1$}{

    \ForEach{$(v_\ell, p_\ell) \in \mathcal{F}$}{
        Sample path-aware query $x_\ell \sim \pi_{\mathrm{plan}}(q, p_\ell)$\;
        
        Select retrieval strategy $r_\ell \sim \pi_{\mathrm{ret}}(x_\ell)$\;
        Retrieve evidence $\mathcal{D}_\ell \leftarrow \mathrm{Retrieve}(x_\ell, r_\ell, \mathcal{D})$\;
        
        Normalize evidence under budget
        $\mathcal{E}_\ell \leftarrow \mathcal{G}_{\mathrm{aug}}(q, \mathcal{D}_\ell, B)$\;
        
        Generate child nodes $\{v_{\ell+1}^{i}\}$ grounded on $\mathcal{E}_\ell$\;
        
        \ForEach{child node $v_{\ell+1}^{i}$}{
            Add edge $(v_\ell \rightarrow v_{\ell+1}^{i})$ to $\mathcal{T}$\;
            Form new path $p_{\ell+1}^{i} = p_\ell \cup \{v_{\ell+1}^{i}\}$\;
            
            Retrieve lightweight evidence $\mathcal{D}_{\ell+1}^{i}$ using $p_{\ell+1}^{i}$\;
            Select structural action
            $a_{\ell+1}^{i} \sim \mathrm{Reflect}(\mathcal{T}, p_{\ell+1}^{i}, \mathcal{D}_{\ell+1}^{i})$\;
            Update tree
            $\mathcal{T} \leftarrow \Phi(\mathcal{T}, v_{\ell+1}^{i}, a_{\ell+1}^{i})$\;

        }
    }
}

\vspace{0.5em}
\textbf{Stage II: Dynamic Adaptive Subtree Selection}

Retrieve global ranking using full tree:
$\mathcal{C} \leftarrow \mathrm{Top}_M(\mathcal{R}(q, \mathcal{T}))$\;

Cluster documents in $\mathcal{C}$ using embeddings:
$\{\mathcal{C}_1,\dots,\mathcal{C}_K\}$\;

\For{$k \leftarrow 1$ \KwTo $K$}{
    Allocate quota
    $n_k = \left\lfloor
    N_{\mathrm{eval}} \cdot \frac{|\mathcal{C}_k|}{|\mathcal{C}|}
    \right\rfloor$\;
    Select $\mathcal{P}_k \leftarrow \mathrm{Top}_{n_k}(\mathcal{C}_k)$\;
}
$\mathcal{P}_{\mathrm{eval}} \leftarrow \bigcup_{k=1}^K \mathcal{P}_k$\;

Annotate $\mathcal{P}_{\mathrm{eval}}$ by LLM; let $\mathcal{P}_{\mathrm{eval}}^+(q)$ denote the relevant subset with size of $\alpha$\;

\ForEach{candidate subtree $\mathcal{T}' \in \mathbb{T}(\mathcal{T})$}{
    Instantiate retrieval $\mathcal{R}(q, \mathcal{T}')|_{\mathcal{P}_{\mathrm{eval}}}$\;
    Estimate Recall@$\,\alpha$ on $\mathcal{P}_{\mathrm{eval}}$\;
}

Select optimal subtree
\[
\mathcal{T}^\star
=
\arg\max_{\mathcal{T}' \in \mathbb{T}(\mathcal{T})}
\mathrm{Recall@}\alpha\bigl(
\mathcal{R}(q, \mathcal{T}')|_{\mathcal{P}_{\mathrm{eval}}},\,
\mathcal{P}_{\mathrm{eval}}^+(q)
\bigr)
\]\;

\Return $\mathcal{T}^\star$\;
\end{algorithm}

\section{Syft News Dataset}
\label{sec:appendix_syft_data}

We evaluate our framework on the Syft News dataset,
which consists of multiple daily snapshots of real-world news streams.
Table~\ref{tab:news_stats} summarizes basic statistics of the dataset
across seven representative dates.

\begin{table}[htbp]
\centering
\caption{Syft News statistics by date. The first line is corpus size while the second line is average relevant documents per query.}
\label{tab:news_stats}
\begin{tabular}{cccccccc}
\toprule
 & 6.11 & 6.12 & 6.13 & 6.14 & 6.15 & 6.29 & 6.30 \\
\midrule
Corpus & 1063 & 1881 & 3960 & 7423 & 14148 & 19320 & 5298 \\
Rel D/Q & 29 & 41 & 95 & 155 & 251 & 369 & 96 \\
\bottomrule
\end{tabular}

\end{table}

\section{Production Recallers in the Syft Online Pipeline}
\label{sec:appendix_recallers}

Syft deploys a large-scale online retrieval pipeline
where multiple production-grade recallers jointly generate
candidate news articles.
We briefly describe each recaller below.

\paragraph{\textbf{Relation-based Recall (RelRec).}}
RelRec retrieves news articles
through explicit relational signals,
such as entity co-occurrence,
knowledge-graph relations,
or historical association patterns,
favoring structurally related content.

\paragraph{\textbf{Decompositional Recall (Decomp).}}
Decomp decomposes complex queries
into simpler sub-queries
and retrieves candidates for each component,
aiming to improve coverage for multi-faceted information needs.

\paragraph{\textbf{Direct Semantic Recall (SemRec).}}
SemRec performs direct dense retrieval
using semantic embeddings
to match queries and documents
based on vector similarity in a shared embedding space.

\paragraph{\textbf{Collaborative Proxy Recall (Proxy).}}
Proxy recall retrieves candidates
via a set of representative proxy queries or documents,
which are collaboratively selected
to approximate diverse user intents
and semantic variations.

\paragraph{\textbf{Web Search Recall (WSRec).}}
WSRec leverages external web search engines
as an auxiliary recall source,
introducing high-quality candidates
from a web-scale retrieval system.
This component complements in-corpus retrieval
by surfacing authoritative and up-to-date content
that may not be well covered
by the internal news index alone.

\section{Construction Cost and Scalability Analysis}
\label{app:cost_analysis}

We further report the absolute construction cost of DynaTree and compare it with representative agentic retrieval baselines under the same topic-level setting. As shown in Table~\ref{tab:construction_cost}, DynaTree incurs higher absolute token and time cost because it constructs a reusable retrieval tree rather than a flat set of semantic expansions. However, its token cost normalized by each generated expansion remains within the same order of magnitude as agentic baselines. Moreover, each DynaTree expansion corresponds to a complete root-to-leaf semantic path, providing richer and reusable subtopic coverage.

The overall cost scales linearly with the number of persistent topic channels. Since Stage~I uses fixed depth and node-count budgets, the per-topic construction cost is bounded and amortized across days. Stage~II further avoids online tree reconstruction and only performs lightweight selection over a constant-size proxy set. 

\begin{table}[htbp]
\centering
\caption{Average construction cost per query theme. 
Token/exp denotes the average token cost per generated semantic expansion.}
\label{tab:construction_cost}
\small
\begin{tabular}{lccc}
\toprule
\textbf{Method} & \textbf{Token/theme} & \textbf{Time/theme} & \textbf{Token/exp} \\
\midrule
ReAct     & 13--16K & 30--40s   & $\sim$1.5K \\
Reflexion & 20--22K & 50--60s   & $\sim$2.2K \\
Self-RAG  & $\sim$15K & 1.5--2min & $\sim$1.5K \\
DynaTree  & $\sim$70K & $\sim$4min & $\sim$3K \\
\bottomrule
\end{tabular}
\end{table}

\section{Shapley-value Attribution Details}
\label{sec:appendix_shapley}

\begin{table}[htbp]
\centering
\small
\setlength{\tabcolsep}{5pt}
\caption{
Details of Shapley-value experiment(7-day average, all values in percentage).
$\checkmark$ indicates enabled component.
}
\label{tab:shapleyyy}
\newcolumntype{C}[1]{>{\centering\arraybackslash}m{#1}}
\begin{tabular}{C{2.2em}C{2.2em}C{2.2em}C{2.2em}|cc}

\toprule
\textbf{P} & \textbf{R} & \textbf{A} & \textbf{F} &
\textbf{Recall@100 (\%)} & \textbf{NDCG@10 (\%)} \\
\midrule
\checkmark & \checkmark & \checkmark & \checkmark & 37.50 & 67.82 \\
\midrule
\checkmark & \checkmark & \checkmark &            & 36.73 & 67.05 \\
\checkmark & \checkmark &            & \checkmark & 37.12 & 67.93 \\
\checkmark &            & \checkmark & \checkmark & 36.62 & 67.48 \\
           & \checkmark & \checkmark & \checkmark & 36.82 & 67.19 \\
\midrule
\checkmark & \checkmark &            &            & 37.47 & 67.53 \\
\checkmark &            & \checkmark &            & 37.52 & 67.27 \\
\checkmark &            &            & \checkmark & 37.18 & 67.53 \\
           & \checkmark & \checkmark &            & 37.08 & 67.48 \\
           & \checkmark &            & \checkmark & 37.12 & 67.55 \\
           &            & \checkmark & \checkmark & 36.70 & 67.63 \\
\midrule
\checkmark &            &            &            & 36.77 & 66.77 \\
           & \checkmark &            &            & 36.53 & 66.45 \\
           &            & \checkmark &            & 36.77 & 67.16 \\
           &            &            & \checkmark & 36.79 & 67.70 \\
\midrule
           &            &            &            & 36.99 & 67.09 \\
\bottomrule
\end{tabular}
\end{table}

This appendix details the formulation and evaluation protocol
for the Shapley-based attribution analysis
used in Section~\ref{exp:ablation1}.
We consider a set of agentic components
$\mathcal{A} = \{P, R, A, F\}$,
corresponding to the Planning, Retrieval, Augmentation,
and Reflection modules in the proposed framework.
For any subset $S \subseteq \mathcal{A}$,
let $f(S)$ denote the retrieval performance
(Recall@100 or NDCG@10)
obtained by enabling exactly the components in $S$.

To capture interaction effects among components,
we exhaustively evaluate all $2^{|\mathcal{A}|} = 16$
possible component combinations
under identical retrieval depth,
semantic expansion budget,
and evaluation protocol.
All results are averaged over seven evaluation days.
Following the standard Shapley-value formulation~\cite{yang2025understandingoptimizingagenticworkflows},
the contribution of component $i \in \mathcal{A}$
is defined as its expected marginal contribution
over all possible coalitions:
\[
\phi_i
=
\sum_{S \subseteq \mathcal{A} \setminus \{i\}}
\frac{|S|!\, (|\mathcal{A}| - |S| - 1)!}{|\mathcal{A}|!}
\big[
f(S \cup \{i\}) - f(S)
\big].
\]
Equivalently, the Shapley value can be interpreted
as the expected marginal gain of component $i$
under a uniformly random ordering of components,
yielding a symmetric and interaction-aware attribution.

Shapley values are computed independently
for Recall@100 and NDCG@10
using the averaged performance $f(S)$.
They reflect the average contribution of each component
under realistic system interactions,
rather than additive or causal importance.
We emphasize that Shapley-based attribution
does not assume component independence
and should not be interpreted
as isolating the intrinsic quality of a module;
instead, it provides a principled summary
of how each component contributes
to overall system performance
in the presence of strong interdependencies.
Table~\ref{tab:shapleyyy} reports the complete results
of all $2^4 = 16$ configurations,
which constitute the full experimental basis
for the attribution analysis.

\section{Uniform vs. Weighted Aggregation}
\label{app:uniform_weighted}

Table~\ref{tab:uniform_weighted_appendix} reports the full seven-day comparison between uniform and softmax-weighted path aggregation. Uniform aggregation consistently provides slightly better or comparable performance, supporting our choice of a simple averaging strategy in the main retrieval scoring function.

\begin{table}[htbp]
\centering
\small
\caption{Per-day comparison between uniform and softmax-weighted path aggregation. Each cell reports Recall@100 / NDCG@10.}
\label{tab:uniform_weighted_appendix}
\begin{tabular}{lcc}
\toprule
\textbf{Date} & \textbf{Uniform} & \textbf{Softmax-weighted} \\
\midrule
6.11 & 0.555 / 0.664 & 0.555 / 0.659 \\
6.12 & 0.579 / 0.709 & 0.529 / 0.695 \\
6.13 & 0.507 / 0.796 & 0.501 / 0.800 \\
6.14 & 0.473 / 0.786 & 0.471 / 0.777 \\
6.15 & 0.408 / 0.791 & 0.394 / 0.793 \\
6.29 & 0.360 / 0.844 & 0.358 / 0.841 \\
6.30 & 0.446 / 0.709 & 0.442 / 0.700 \\
\midrule
\textbf{Avg} & \textbf{0.475 / 0.757} & 0.464 / 0.752 \\
\bottomrule
\end{tabular}
\end{table}

\section{Case Study: Robustness to Emerging Entities}
\label{sec:appendix_casestudy}

Although the retrieval tree is fixed after Stage~I, DynaTree does not rely on exact lexical overlap with future news entities. We build the tree on the June~29 snapshot and inspect retrieved articles on June~30. As shown in Table~\ref{tab:emerging_entities}, DynaTree retrieves articles containing entities absent from the original tree because higher-level semantic paths still capture their topical context. This indicates that the tree acts as a reusable semantic routing structure while document-level retrieval remains open to new surface realizations. For disruptive zero-day topics, the system falls back to existing low-latency recallers and asynchronously applies local Add/Revise operations via Reflection Agent, avoiding full online reconstruction.

\begin{table}[htbp]
\centering
\small
\caption{Examples of emerging entities retrieved on June~30 using a tree constructed on June~29.}
\label{tab:emerging_entities}
\begin{tabular}{ll}
\toprule
\textbf{Branch} & \textbf{New entities in retrieved news} \\
\midrule
NAS OS      & Feiniu, fnOS \\
Wearables   & Soundcore P41i, PITAKA \\
EV Battery  & Harrier PHV \\
\bottomrule
\end{tabular}
\end{table}


\end{document}